\documentclass[intlimits,twoside,a4paper]{article}

\usepackage{amsmath,amssymb}
\usepackage{graphicx}
\usepackage{subfig}
\usepackage[T2A]{fontenc}
\usepackage[cp1251]{inputenc}

\usepackage[eqsecnum]{cmpj3}

\issue{2021}{24}{1}{13703}
\doinumber{10.5488/CMP.24.13703}

\title[Prediction of electronic and half metallic properties..,]%
{Prediction of electronic and half metallic properties of Mn$_2$YSn (Y = Mo, Nb, Zr) Heusler alloys%
}
\author[S. Zeffane \textsl{et al.}]{S. Zeffane\refaddr{label1,label2}, M. Sayah\refaddr{label1,label2}, F. Dahmane\refaddr{label1,label3}, M. Mokhtari\refaddr{label1,label2}, L. Zekri\refaddr{label2}, R. Khenata\refaddr{label3}, N.~Zekri\refaddr{label2}}
\addresses{
\addr{label1} D\'epartement de SM, Institut des Sciences et des Technologies, Centre Universitaire de Tissemsilt, 38000 Tissemsilt, Alg\'erie 
\addr{label2} Universit\'e des Sciences et de la Technologie d'Oran Mohamed Boudiaf, USTO-MB, LEPM, BP 1505, El M' Naouar, 31000 Oran, Algeria 
\addr{label3} Laboratoire de Physique Quantique et de Mod\'elisation Math\'ematique (LPQ3M), D\'epartement de Technologie, Universit\'e de Mascara, 29000 Mascara, Alg\'erie 
 }

\date{Received April 23, 2020, in final form September 5, 2020}
\begin{document}

\maketitle

\begin{abstract}

We investigate the structural, electronic and magnetic properties of the full Heusler compounds Mn$_2$YSn (Y~=~Mo, Nb, Zr) by first- principles  density functional theory using the generalized gradient approximation. It is found that the calculated lattice constants are in good agreement with the theoretical values. We observe that the Cu$_2$MnAl-type structure is more stable than the Hg$_2$CuTi type. The calculated total magnetic moments of \linebreak   Mn$_2$NbSn and Mn$_2$ZrSn are 1 $\mu_{\textrm{B}}$ and 2 $\mu_{\textrm{B}}$ at the equilibrium lattice constant of 6.18 \AA ~and 6.31 \AA, respectively, for the Cu$_2$MnAl-type structure. Mn$_2$MoSn have a metallic character in both Hg$_2$CuTi and Cu$_2$MnAl type structures. The total spin magnetic moment obeys the Slater-Pauling rule. Half-metal exhibits $100\%$ spin polarization at the Fermi level. Thus, these alloys  are promising magnetic candidates in spintronic devices. 


\end{abstract}

\section{Introduction}
To ameliorate the performance of spintronic devices is indispensable for the advance of modern technology \cite{1}. Half-metallic ferromagnets (HMFs) are interesting spin-polarized materials and, thus these are ideal for the application in spintronic devices \cite {2}. Half-metal magnets have got broad attentions.  They are utilized in the manufacturing of electronic gadgets because of their wide band gap in minority spins, magnetic random access memory effect, high data processing rate and low consumption of electric power and gradually increasing density \cite{33, 44}

 Numerous half-metallic ferromagnets have been predicted and verified experimentally since NiMnSn was predicted in1983 by De Groot et al. \cite{3}. Ferromagnetic materials display diverse electronic properties in the spin up and down bands, with metallic properties in one spin band and insulator or semiconductor properties in another, thus leading to $100\%$ spin polarization at the Fermi level    \cite{5} \cite{6} \cite{77}. Heusler alloys are a class of inter-metallic compounds, simple structures and unique properties \cite{7}. In 1903, a German scientist Heusler found that the atoms in the alloy Cu$_2$MnAl were non-magnetic (NM), but the alloys showed an adjustable magnetism through heat treatment and chemical components. During the past few decades, Heusler alloys have been favorable candidates for multifunctional materials because of their numerous excellent properties, such as: Magnetocaloric effect \cite{8} \cite{9}, giant magnetoresistance \cite{10}, magnetic field-drive shape memory effects \cite{11}, half-metallicity \cite{12}, Hall effects \cite{13}. In addition, some Heusler compounds  exhibit excellent thermoelectric properties \cite{14, 155}.

Furthermore, a very interesting class of Heusler alloys that has received  considerable  theoretical  studies  is  the  HM,  Mn$_2$YZ. These materials  are  much  more  favorable  than  their  ferromagnetic counterparts in magneto-electronic applications \cite{116}. One important application of Mn$_2$YZ Heusler alloys is spintronic materials. Many  Mn$_2$-based Heusler alloys have been reported to be half-metals or spin gapless semiconductors (SGSs) such as Mn$_2$CoAl inverse Heusler alloy under pressure \cite{117}, Mn$_2$CoZ  (Z = Al, Ga, In, Si, Ge, Sn, Sb) \cite{118}, Mn$_2$VZ (Z = Al, Ga, In, Si, Ge, Sn)  \cite{119}, Mn$_2$CoAl \cite{120}.

 In this paper, we present an investigation on the structural, electronic, magnetic properties and halfmetallic behavior of Mn$_2$YSn (Y= Mo, Nb, Zr). This paper is structured as follows: in section~\ref{sec2}, we briefly describe the computational method used in this work, Results and discussions of our study are present in section~\ref{sec3}. Finally, a summary of the work is given in section~\ref{sec4}.

\section{Computational method}\label{sec2}

In order to calculate the electronic, structure and magnetic properties, we employed the FP-LAPW method in the framework of the density functional theory (DFT) \cite{15} as implemented in the WIEN2k code~\cite{16} .In this method, the space is divided into non-overlapping muffin-tin (MT)	sphere separated by an interstitial region. The generalized approximation proposed by Perdew-Burke-Ernzerhof was used for the exchange correlation potential \cite{17}. Spin polarized calculations were performed with both spin-up and spin-down states. The maximum value of angular momentum $ L_\text{max}= 10$ for the wave function expansion inside the muffin tin sphere. The convergence of the basis was controlled by cut-off of $ K_\text{ max}= 8.0 /R_\text{MT}$ where $R_\text{MT}$ is the muffin tin sphere radius and $K_\text{max}$ is the largest reciprocal lattice vector used in the plane wave expansion within the interstitial region.  
The cutoff energy, which defines the separation of valence and core states, was chosen as $-6.0$~Ry. A mesh of 64 special k-points was made in the irreducible wedge of the Brillouin zone. 
In the interstitial region, the charge density and the potential were expanded as a Fourier series with wave vectors up to $G_\text{max}=12$ a.u.$^{-1}$. A convergence norm for self-consistent field calculations was chosen in such a way that the difference in the energy between two successive iterations did not exceed $10 ^{-4}$ Ry.

The radii $R_\text{MT}$ of the muffin-tin are selected to be as large as possible under the condition that the spheres do not overlap.
The electronics configurations for atoms in Mn$_2$YSn (Y = Mo, Nb, Zr) are:  
Mn: $[Ar]4s^2 3d^5$, Mo: $[Kr]4s^1 4d^5$, Nb: $[Kr]5s^1 4d^4$, Zr: $[Kr]5s^24d^2$, Sn: $[Kr]5s^2 4d^{10}5p^2$.

\section{Results and discussion}\label{sec3}

\begin{figure}[!t]%
    \centering
    \subfloat[]{{\includegraphics[width=7cm]{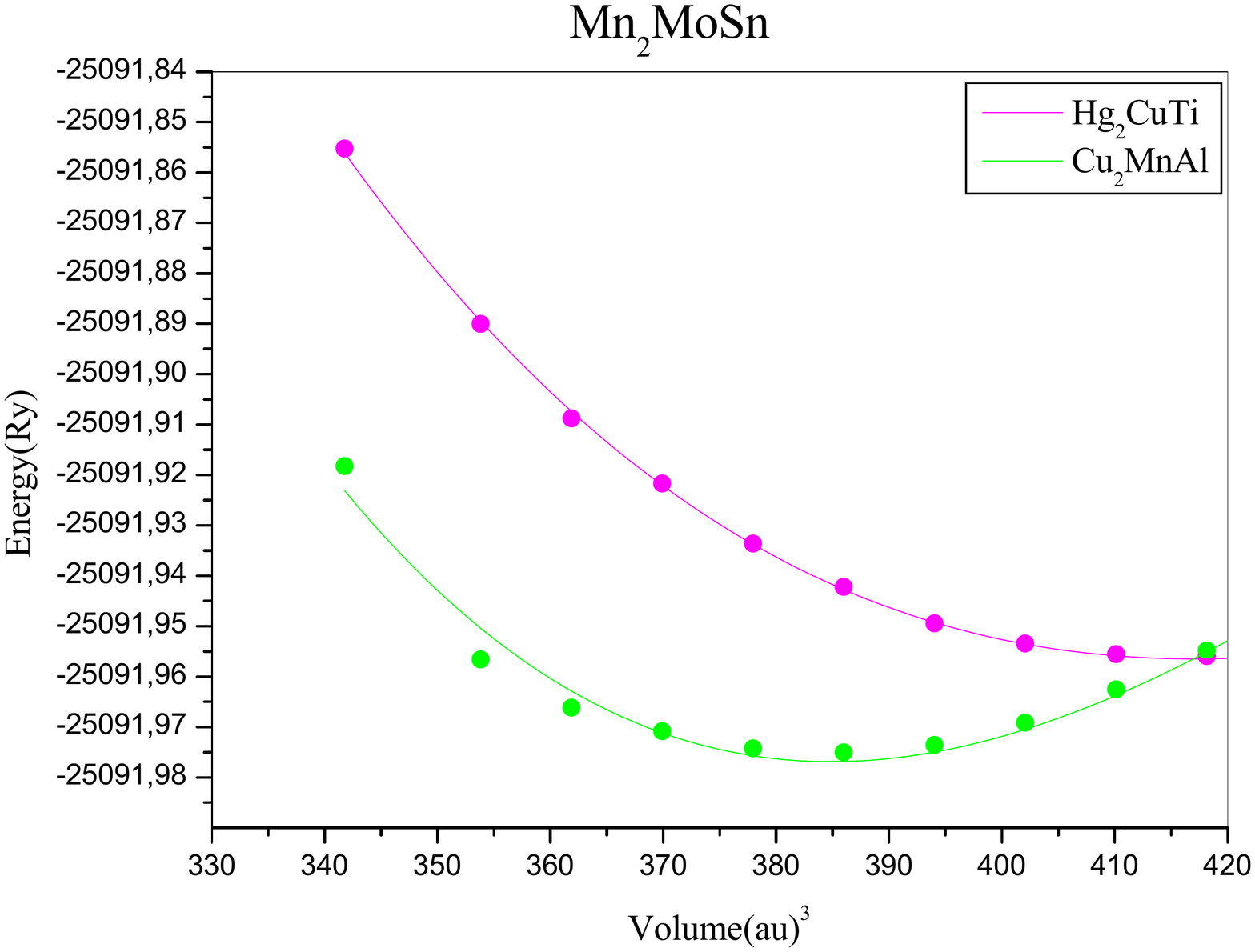} }}%
    \qquad
    \subfloat[]{{\includegraphics[width=7cm]{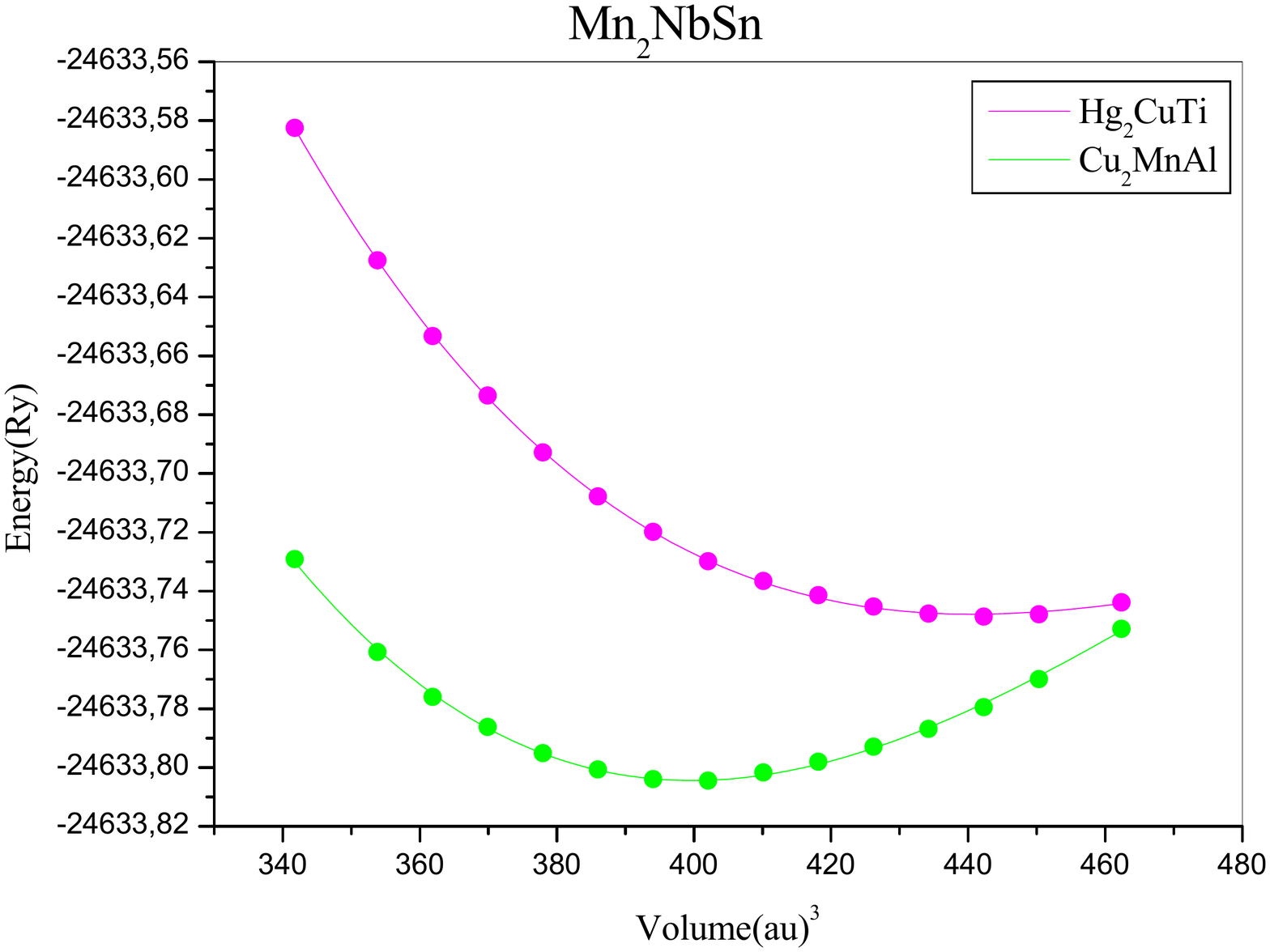} }}%
 \qquad
    \subfloat[]{{\includegraphics[width=7cm]{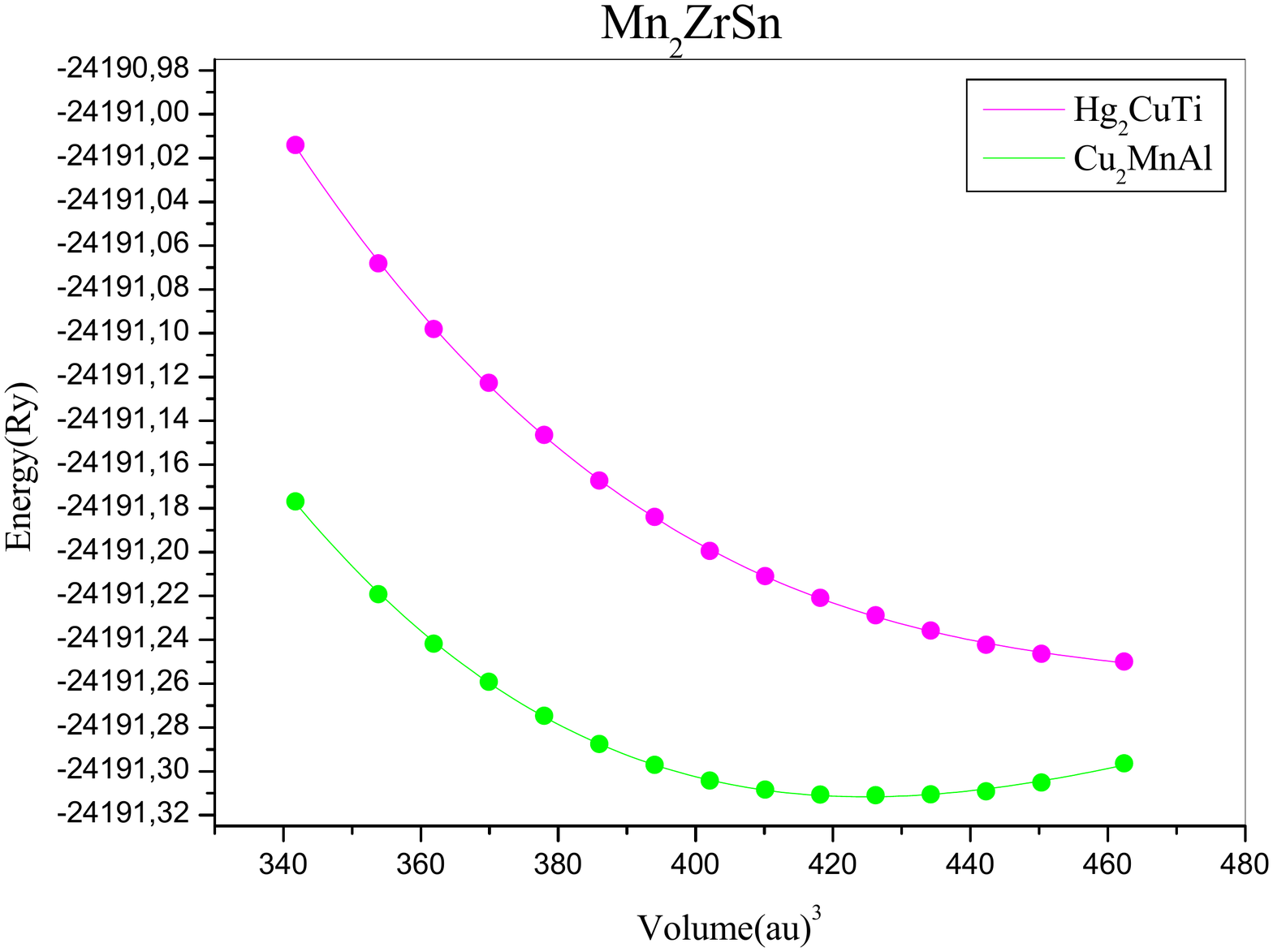} }}%
    \caption{(Colour online) Total energy versus lattice constant for both structure Hg$_2$CuTi and for Mn$_2$YSN (Y = Mo, Nb, Zr).}%
    \label{fig:example}%
\end{figure}

In this subsection, we present the results of the geometrical structure of the Mn$_2$YSn (Y = Mo, Nb, Zr) Heusler alloys as well as the lattice parameters and bulk modulus. The general stoichiometric composition of the full-Heusler alloys is X$_2$YZ, where X and Y are different transition elements, while Z refers to the main group element. The high ordered structure is a very important factor for the electronic, magnetic properties of Heusler compounds. There are two possible atomic orderings in Mn$_2$YSn with (Y = Mo, Nb, Zr): the first one is L21 (``regular cubic phase'' prototype Cu$_2$MnAl) in which the two Mn atoms occupy A(0,0,0) and C(1/2, 1/2, 1/2) positions, and Y, Sn atoms occupy B (1/4,1/4,1/4) and D(3/4, 3/4, 3/4). In the case of the Cu$_2$MnAl type L21 structure, the sequence of the atoms occupying the four sites of the unit cell is X-Y-X-Z. The second one is XA (``inverted cubic phase'' prototype Hg$_2$CuAl, in which the tow Mn atoms occupy A(0,0,0) and B (1/4,1/4,1/4) positions, and Y, Sn atoms occupy C (1/2,1/2,1/2) and D (3/4,3/4,3/4) positions, respectively, the sequence of the atoms  is X-X-Y-Z. The important difference between these two structures is the inter-exchange between the C site atom and the B-site atom. Both structures may be indistinguishable by X-ray diffraction and much care should be taken in the structural analysis, because both have the general FCC like symmetry \cite{18} \cite{19}. To determine the ground state properties of Mn$_2$YSn (Y = Mo, Nb, Zr), the calculation results of total energy versus lattice constant for both Hg$_2$CuTi and Cu$_2$MnAl structures are plotted in figure~\ref{fig:example}. 
The variation of total energy with the volume is fitted to Murnaghan equation of state \cite{20} to obtain the equilibrium lattice constant a \AA, the bulk modulus $B$ (GPa), the derivative of the bulk with respect to the modulus $B$'.

\begin{equation}
E(V)=E_{0}(V)+\frac{BV}{B'(B'-1)}\left[ B\left( 1-\frac{V_{0}}{V}\right) +\left( \frac{V_{0}}{V}\right) ^{B'}-1\right], 
\label{moneq}
\end{equation}
where is $E_0$ the minimum energy at $T= 0$ K, $B$ is the bulk modulus, $B$' is the bulk modulus derivative and $V_0$ is the equilibrium volume.
Clearly, for Mn$_2$YSn (Y = Mo, Nb, Zr) compounds, the regular cubic phase (prototype Cu$_2$MnAl) Heusler structure is  more stable than the inverted cubic phase (prototype Hg$_2$CuAl).
The results are listed in table~\ref{tbl-smp1}. The calculated lattice constant of Mn$_2$YSn (Y = Mo, Nb, Zr) is in good agreement with the previously theoretically optimized lattice constants reported by other researchers.
According to Luo and al \cite{21}, the site preference of the X and Y atoms is strongly influenced by the number of their electrons. Those elements with more electrons prefer to occupy the A and C sites, and those with fewer electrons tend to occupy the $B$ sites. In the case of Mn$_2$YSn (Y = Mo, Nb, Zr), we have the nuclear charge of X (Mn atom)  larger than Y (Y = Mo, Nb, Zr), so the Cu$_2$MnAl structure will be observed. Similar results are found by Kervan and al \cite{22} for Mn$_2$NbAl, and Anjami and al \cite{12} for Mn$_2$ZrX (X = Ge, Si).
We discuss the phase stability of Mn$_2$YSn (Y = Mo, Nb, Zr) based on the formation energy ($\Delta E_\text{F}$). This can help to predict whether these alloys can be prepared experimentally. Here, the formation energy $\Delta E_\text{F}$ is calculated by comparing the total energies of the Mn$_2$YSN (Y = Mo, Nb, Zr) Heusler alloys with the sum of the total energies of the constituting elements.
The formation energy of the Mn$_2$YSn (Y = Mo, Nb, Zr) materials is computed following the expression given below

\begin{equation}
\Delta E_\text{F}=E^{\text{Total}}_{\text{Mn}_2\text{YSn}}-\left[ 2 E^{\text{bulk}}_{\text{Mn}}+E^{\text{bulk}}_{\text{Y}}+E^{\text{bulk}}_{\text{Sn}}\right] ,
\label{moneq}
\end{equation}
where $ E^{\text{Total}}_{\text{Mn}_2\text{YSn}}$  is the first-principles computed results of the total energy at equilibrium for the Mn$_2$YSn (Y = Mo, Nb, Zr) full Heusler alloys, and  $E^{\text{bulk}}_{\text{Mn}}$, $E^{\text{bulk}}_{\text{Y}}$, $E^{\text{bulk}}_{\text{Sn}}$  represent total energy per atom for Mn, Y, and Sn elements in the bulk form, respectively. The negative values of the formation energy specify that Mn$_2$YSn (Y = Mo, Nb, Zr) full Heusler alloys are chemically stable, and these materials can be synthesized experimentally. The computed results of the $\Delta E_f$ for the Cu$_2$MnAl type structures are found to be more negative than these of the Hg$_2$CuTi type structures, endorsing that Cu$_2$MnAl type structures are more stable  compared to the Hg$_2$CuTi type ones.

 \begin{table}[!t]
\caption{Calculated equilibrium lattice constant $a$ (\AA), the bulk modulus $B$ (GPa), the minimum energy (Ry) and the formation energy $E_\text{F}$ (Ry)   of Mn$_2$YSn (Y = Mo, Nb, Zr) Heusler compounds.}
\label{tbl-smp1}
\vspace{2ex}
\begin{center}
\renewcommand{\arraystretch}{0}
\begin{tabular}{|c|c||c|c|c|c|c||}
\hline
      & &$a$ (\AA)&$B$ (GPa)&$B'$&Energy (Ry)&$E_\text{F}$ (Ry)\strut\\
\hline
\rule{0pt}{2pt}&&&&&\\
\hline
\raisebox{-1.7ex}[0pt][0pt]{Mn$_2$MoSn} 
 &  Hg$_2$CuTi&6.2825&133.9353&5.2946&$- 25091.956309$&$-1.257851$\strut\\ 
\cline{2-7}
 & Cu$_2$MnAl&6.1033&264.5693&6.9898&$-25091.976771$&$-1.278313$\strut\\
\hline
\raisebox{-1.7ex}[0pt][0pt]{Mn$_2$NbSn}   
& Hg$_2$CuTie&6.3946&138.8278&4.0864&$- 24633.748350$&$-1.296929$\strut\\
\cline{2-7}
& Cu$_2$MnAl&6.1872&195.5253&4.4311&$-24633.803953$&$-1.322501$\strut\\
\hline
\raisebox{-1.7ex}[0pt][0pt]{Mn$_2$ZrSn}
 &Hg$_2$CuTi&6.5743&98.2694&4.2436&$-24191.251440$&$- 1.149113$\strut\\ 
\cline{2-7}      
& Cu$_2$MnAl&6.3195&162.0852&4.0633&$-24191.311742$&$-1.192513$\strut\\
\hline
\end{tabular}
\renewcommand{\arraystretch}{1}
\end{center}
\end{table}

\begin{figure}[!t]
\begin{minipage}[c]{.45\linewidth}
\begin{center}
\includegraphics[scale=0.7]{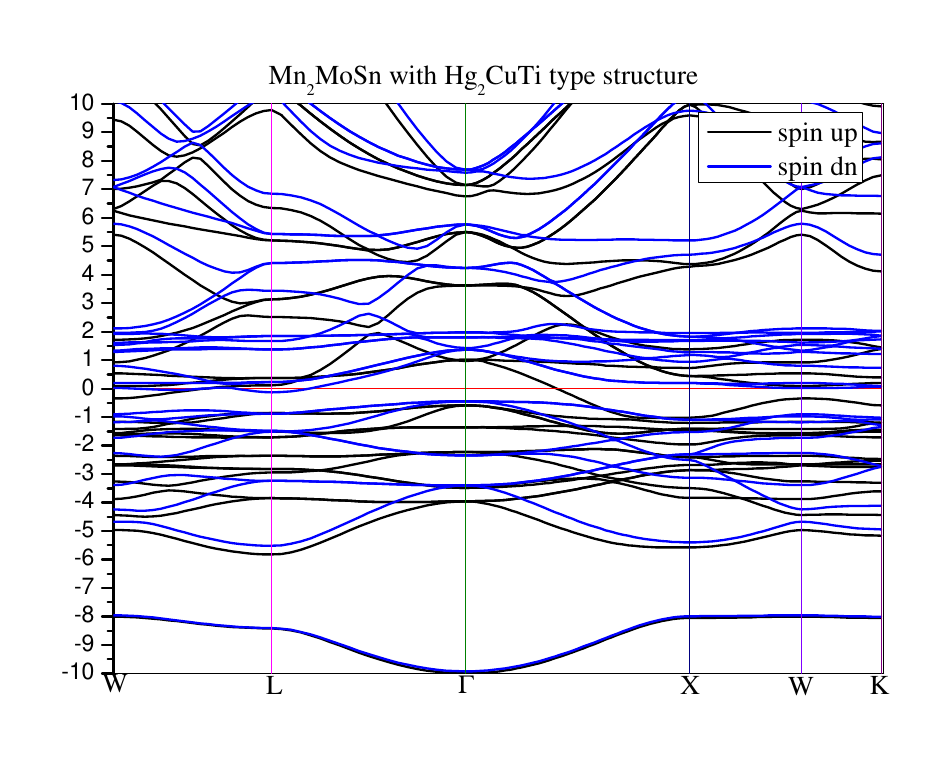}
\end{center}
\end{minipage}
\hfill
\begin{minipage}[c]{.45\linewidth}
\begin{center}
\includegraphics[scale=0.7]{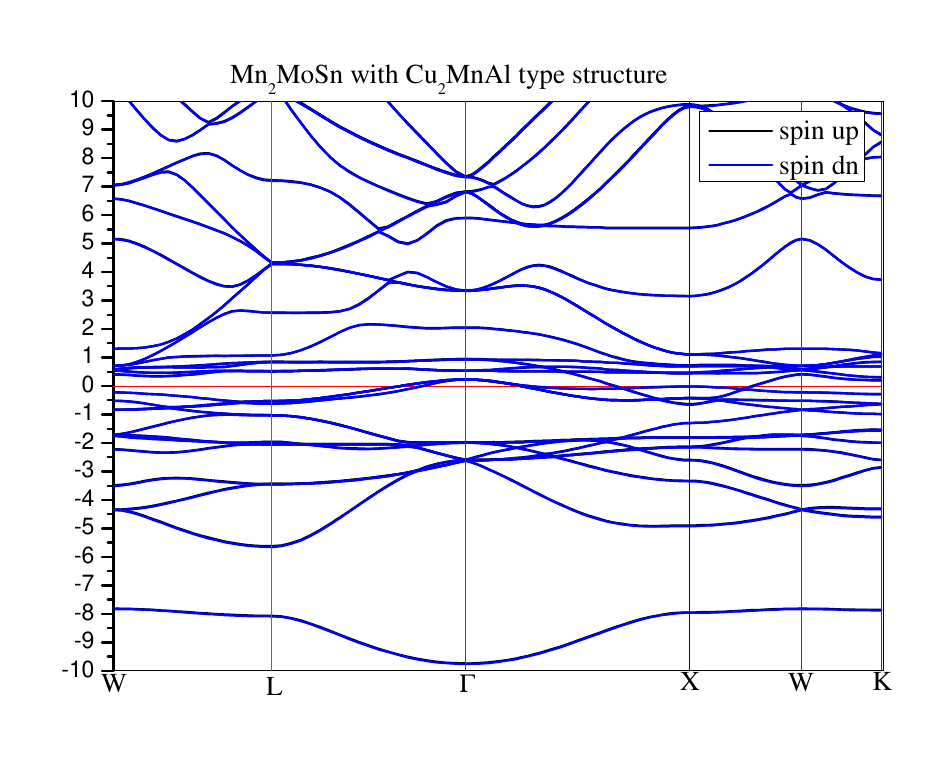}
\end{center}
\end{minipage}
\begin{minipage}[c]{.45\linewidth}
\begin{center}
\includegraphics[scale=0.7]{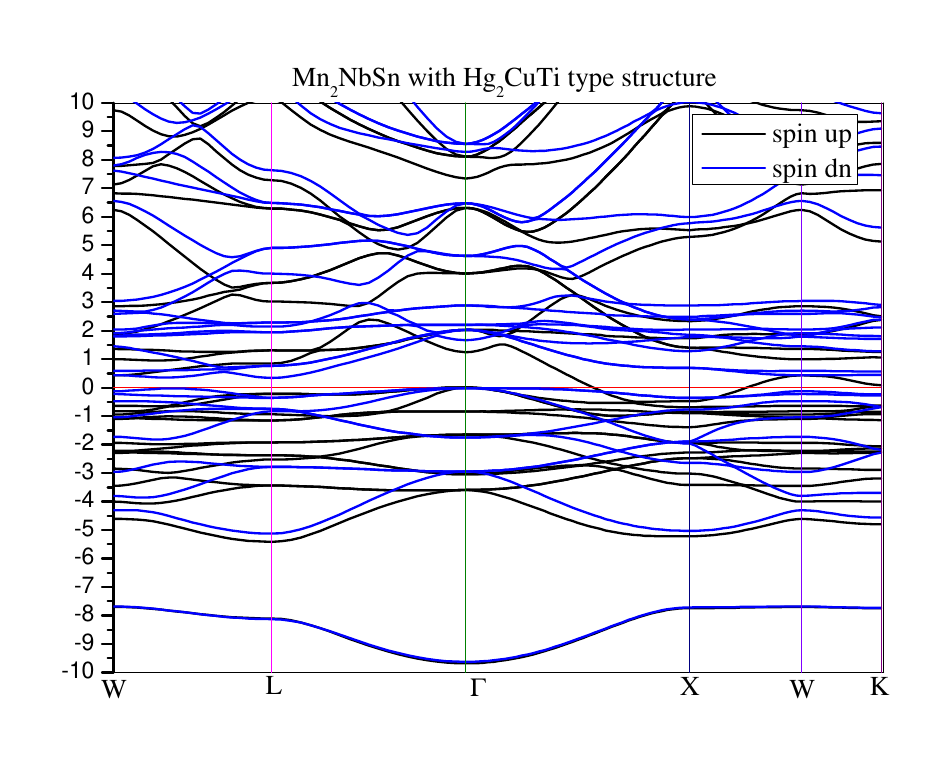}
\end{center}
\end{minipage}
\hfill
\begin{minipage}[c]{.45\linewidth}
\begin{center}
\includegraphics[scale=0.7]{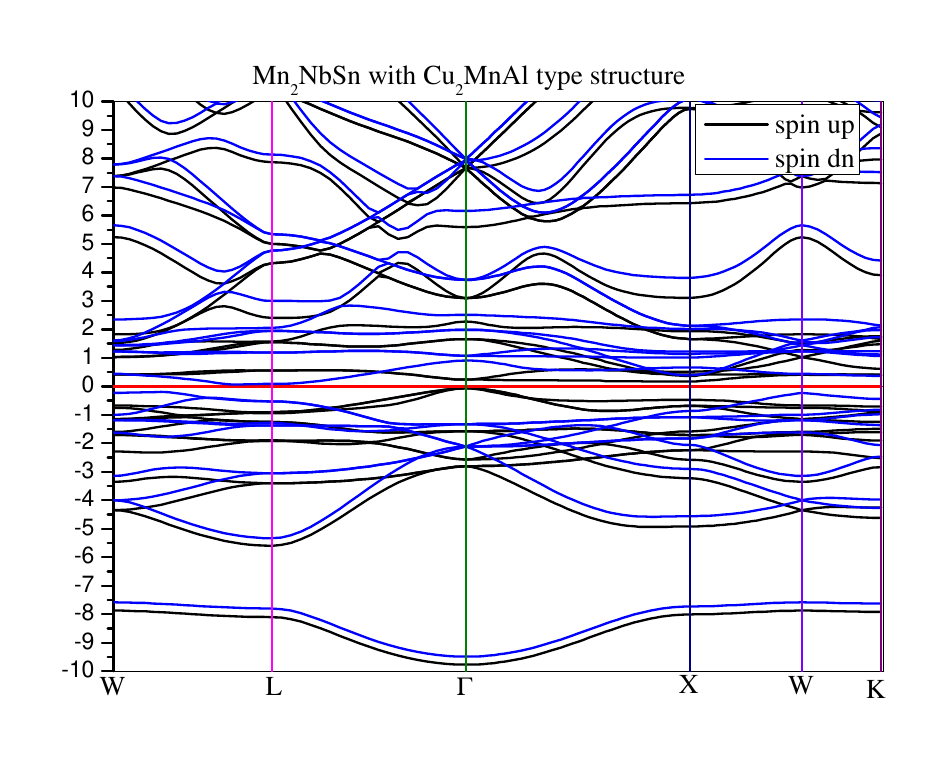}
\end{center}
\end{minipage}
\begin{minipage}[c]{.45\linewidth}
\begin{center}
\includegraphics[scale=0.7]{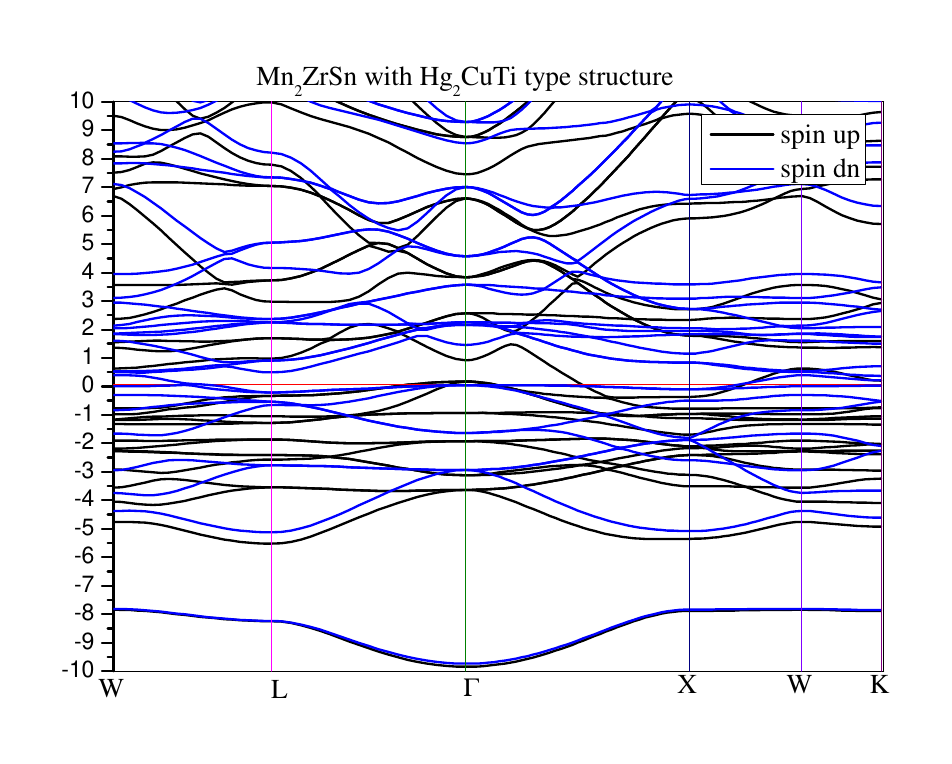}
\end{center}
\end{minipage}
\hfill
\begin{minipage}[c]{.45\linewidth}
\begin{center}
\includegraphics[scale=0.7]{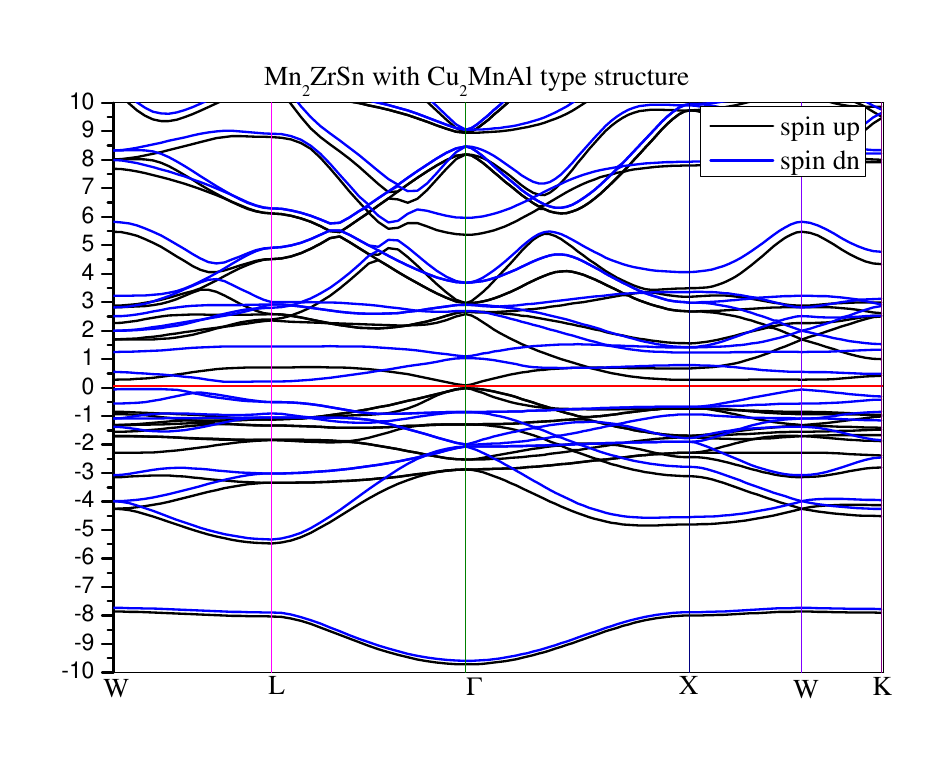}
\end{center}
\end{minipage}
\caption{(Colour online) Band structure for Mn$_2$YSn (Y = Mo, Nb, Zr)  for both structure Hg$_2$CuTi and Cu$_2$MnAl. }
\label{fig2}
\end{figure}

\begin{figure}[!t]
\begin{minipage}[c]{.45\linewidth}
\begin{center}
\includegraphics[scale=0.24]{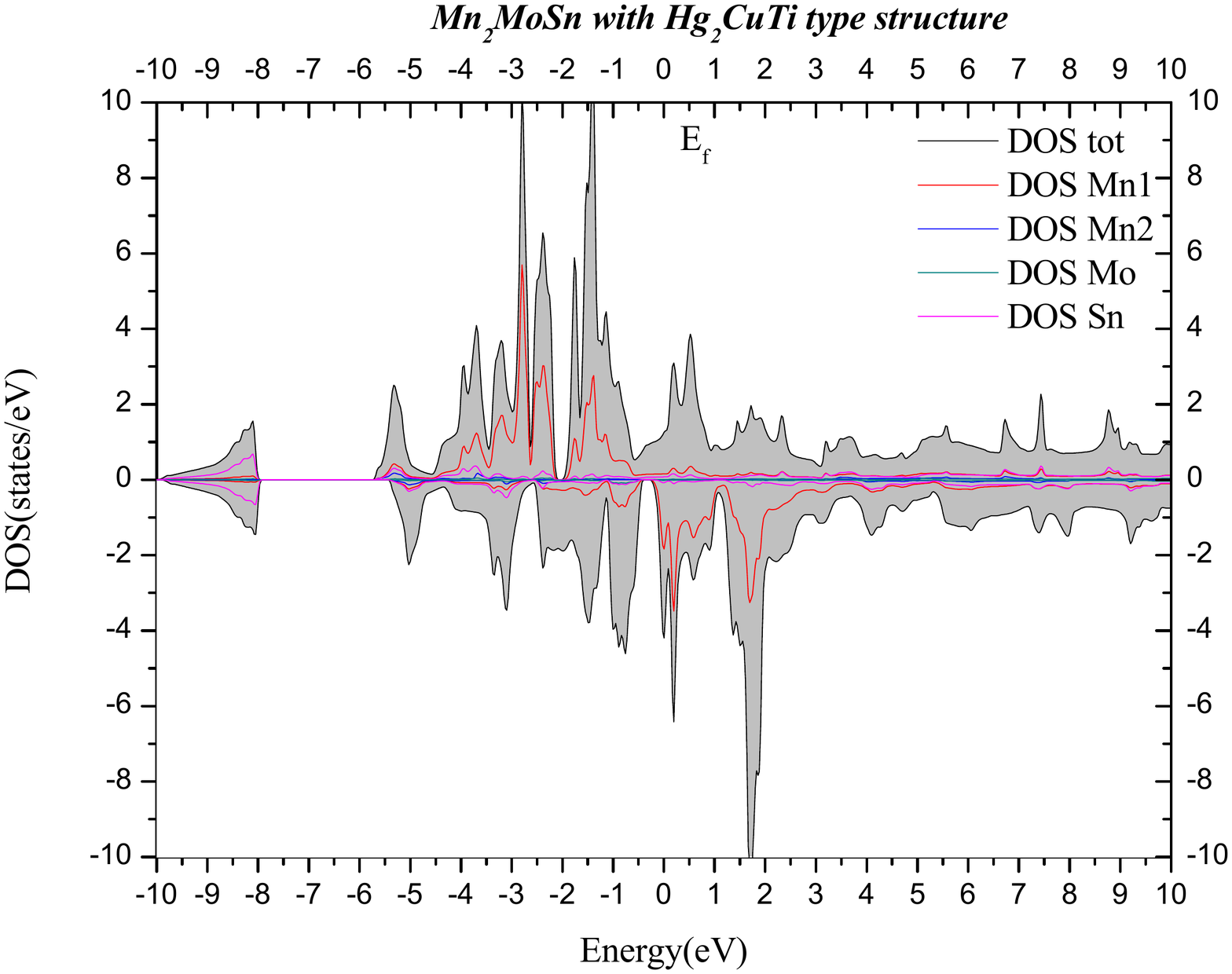}
\end{center}
\end{minipage}
\hfill
\begin{minipage}[c]{.45\linewidth}
\begin{center}
\includegraphics[scale=0.24]{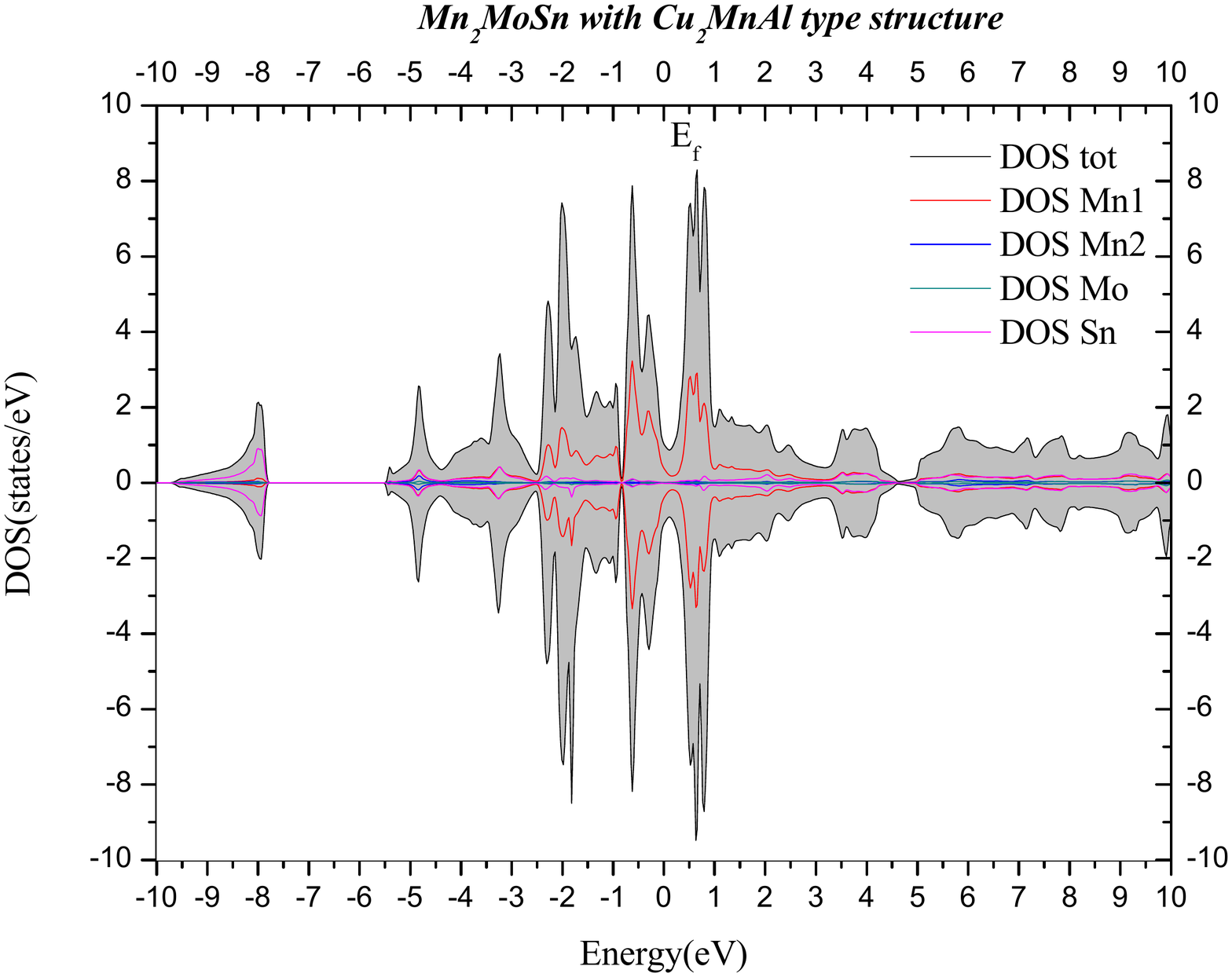}
\end{center}
\end{minipage}
\begin{minipage}[c]{.45\linewidth}
\begin{center}
\includegraphics[scale=0.24]{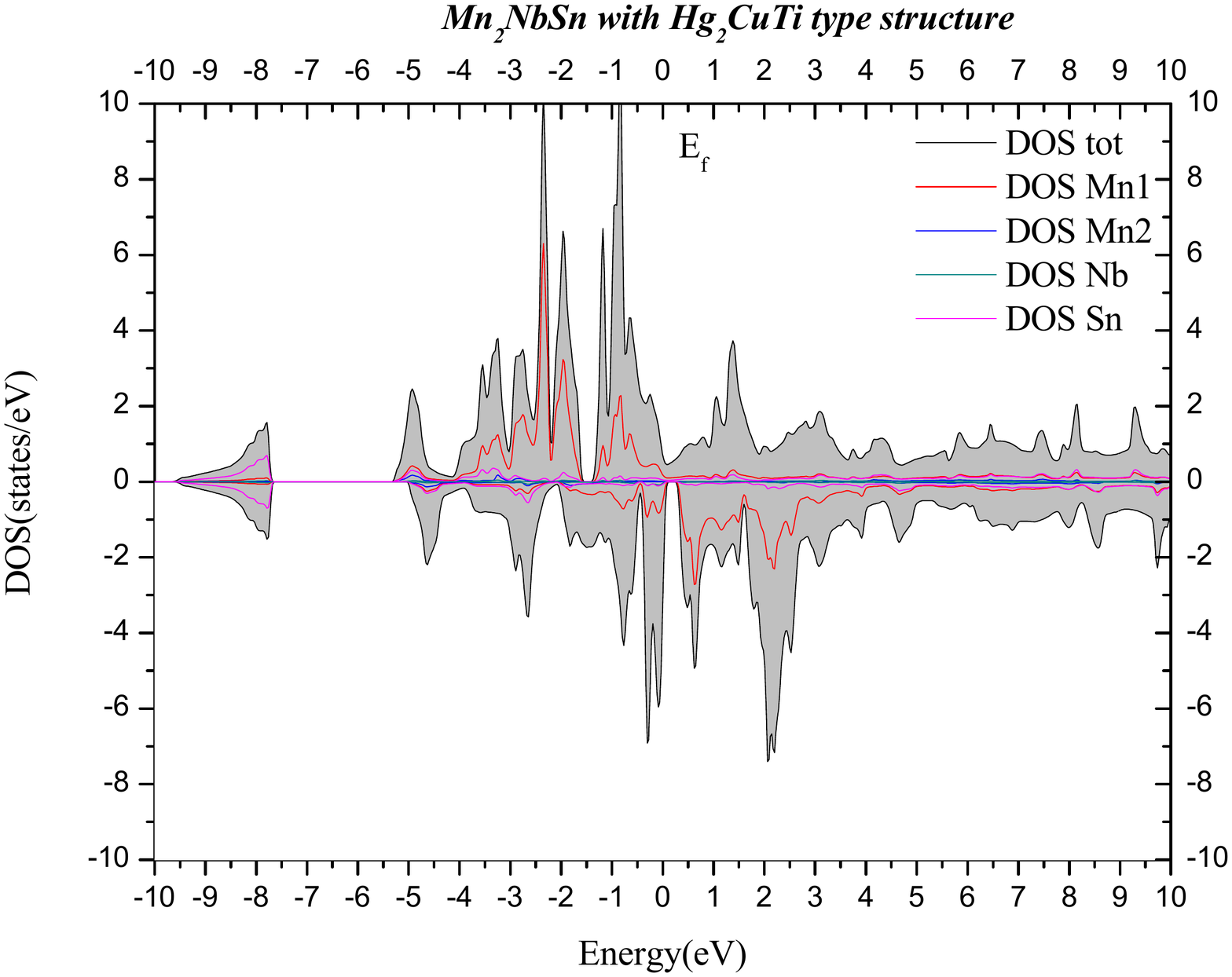}
\end{center}
\end{minipage}
\hfill
\begin{minipage}[c]{.45\linewidth}
\begin{center}
\includegraphics[scale=0.25]{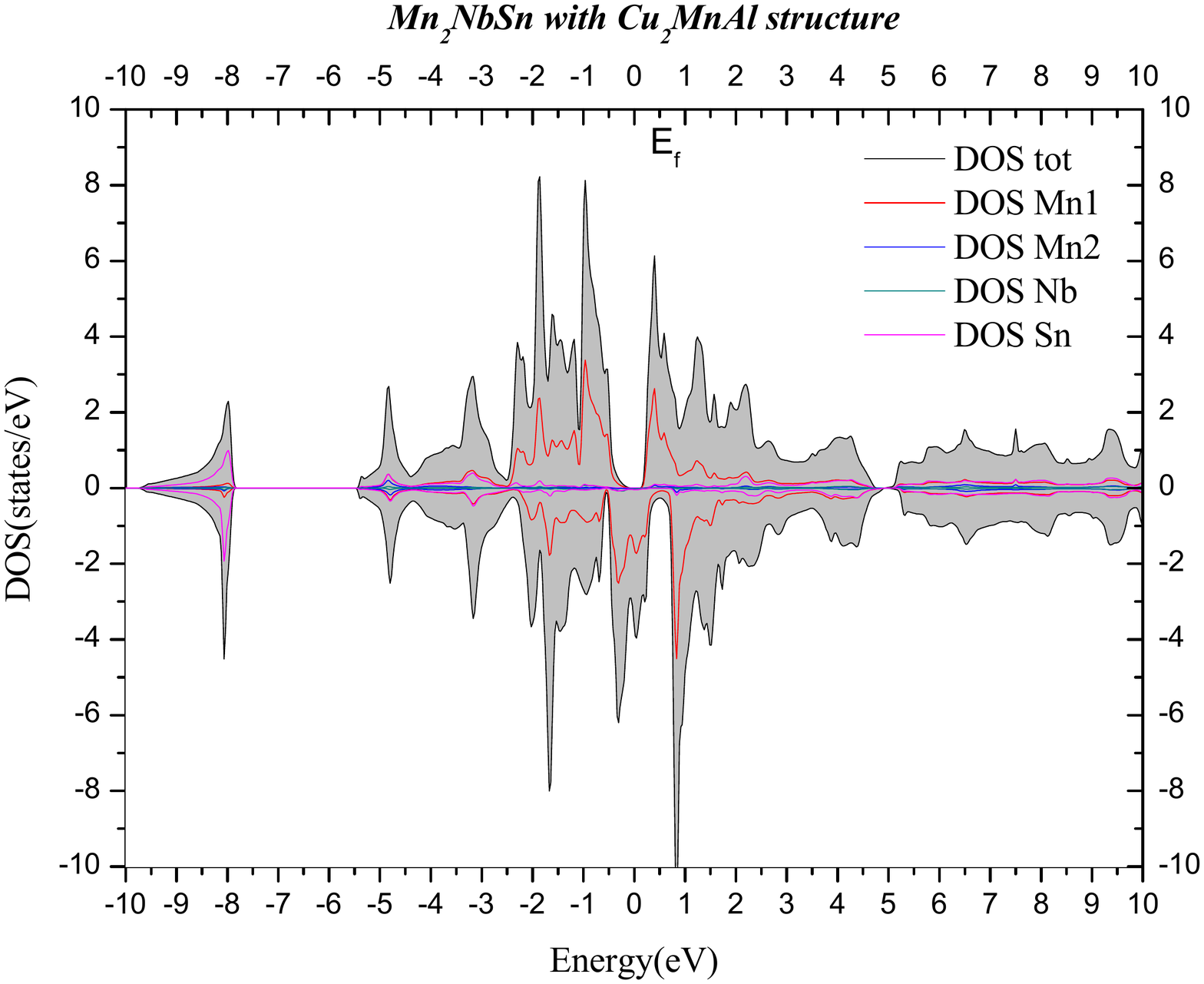}
\end{center}
\end{minipage}
\begin{minipage}[c]{.45\linewidth}
\begin{center}
\includegraphics[scale=0.24]{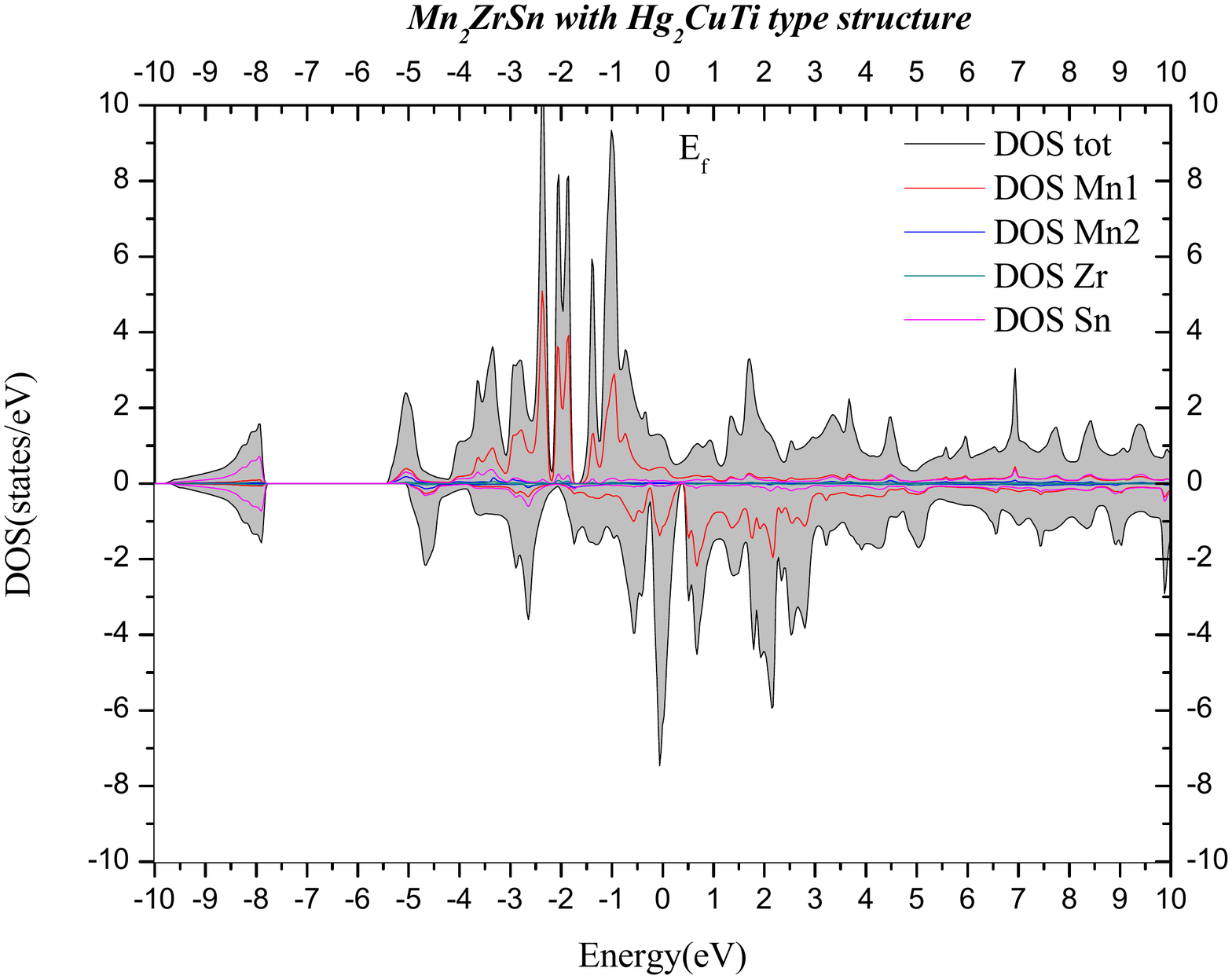}
\end{center}
\end{minipage}
\hfill
\begin{minipage}[c]{.45\linewidth}
\begin{center}
\includegraphics[scale=0.24]{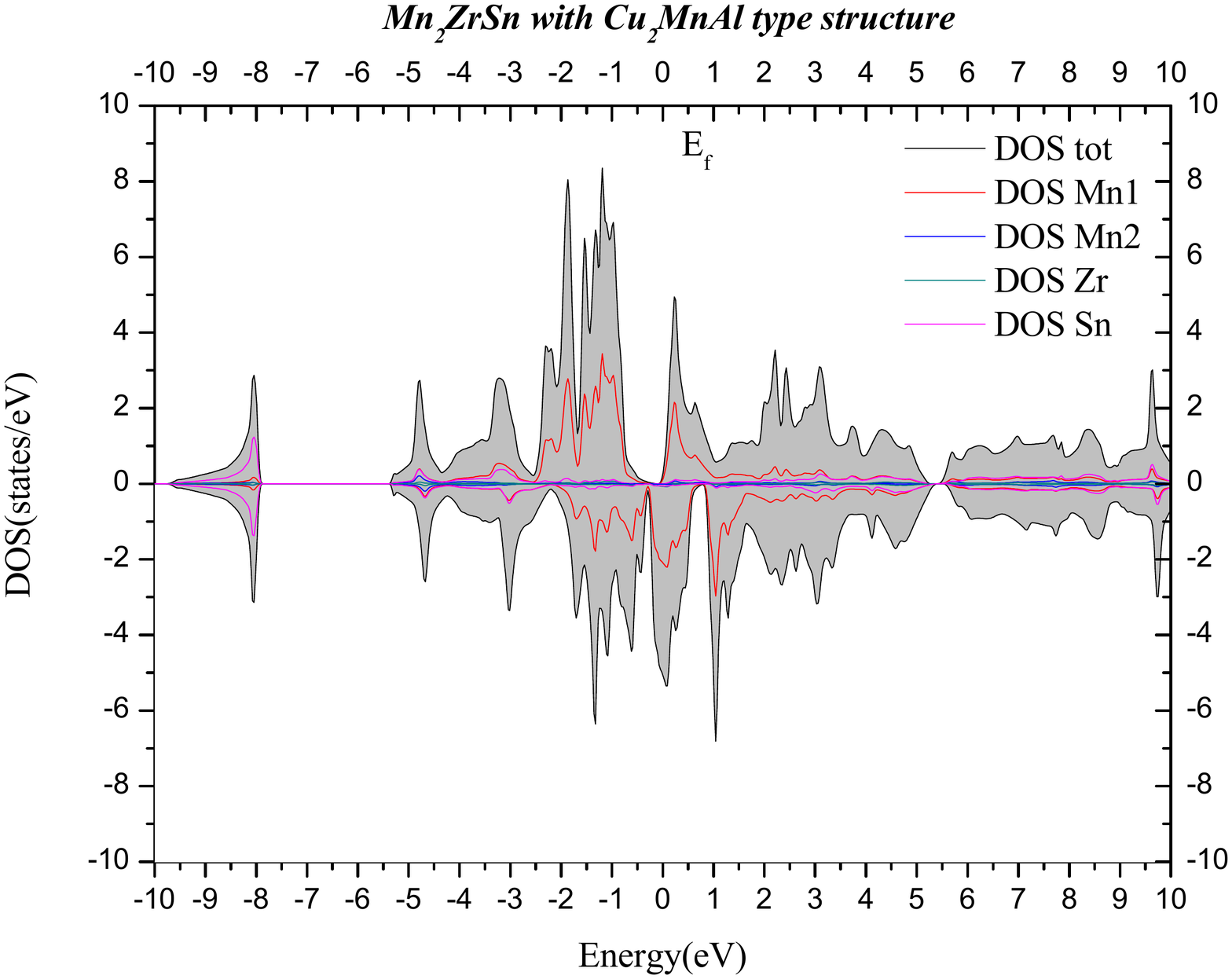}
\end{center}
\end{minipage}
\caption{(Colour online) Total and partial density 
of states for Mn$_2$YSn (Y = Mo, Nb, Zr) for both structure Hg$_2$CuTi and Cu$_2$MnAl. }
\label{fig3}
\end{figure}

The gap in half-metallic Heusler alloys exist in one state, whereas in the opposite spin state, $E_f$   cuts though the bands. The d-band is mostly responsible for the position of the Fermi level lying in it. The role of transition metals d-states is very essential in the explanation of spin-polarized electronic band structures; the density of states of one spin state has a peak at $E_f$ while in another spin state, the density of the state is zero around the $E_f$  \cite{23}.
The conduction electrons are thus $100\%$ spin-polarized, and it is useful to define the electron spin polarization at the Fermi energy of a material where the spin polarization at $E_f$  is given by equation. 
\begin{equation}
P=\frac{N_{\uparrow} (E_f)-  N_{\downarrow} (E_f) }{N_{\uparrow} (E_f) + N_{\downarrow}(E_f)},
\label{moneq}
\end{equation}
where $N_{\uparrow} (E_f)$ and $ N_{\downarrow} (E_f)$ are the spin dependent densities of states at $E_\text{F}$, the  $\uparrow$ and $\downarrow$  assign states of the opposite spins, that are the majority and minority states, respectively.
With the aim to  profoundly understand the electron structures of  Mn$_2$YSn (Y = Mo, Nb, Zr), we present the energy bands along high symmetry directions in the Brillouin zone, the total density of states (TDOS) and partial density of states (PDOS) plots for  Mn$_2$YSn (Y = Mo, Nb, Zr)  in figure~\ref{fig2}. PDOS plots are plotted to see the contributions from various atomic states near $E_\text{F}$. We treat the spin-up channel as positive and the spin-down channel as negative for better comparison.
The calculated spin-polarized total densities of states (DOS) and atom projected DOS of theMn$_2$MoSn compound for the Cu$_2$MnAl type structure and the Hg$_2$CuTi type structure are presented at their optimized equilibrium lattice constants in figure~\ref{fig2}. We can see in Cu$_2$MnAl structure that the spin up and spin down are symmetric and the band structures are identical explaining the non-magnetic behavior of this alloy. For Hg$_2$CuTi type structure Mn$_2$MoSn is metallic with intersection of the band structure with Fermi level in spin up and spin down.
For the  Mn$_2$ZrSn, the minority spin channel (spin down) has intersection with Fermi level, so it has a metallic character, while the majority spin channel (spin up) has a gap at Fermi level. Consequently, it shows a semiconductor behavior. As presented in figure~\ref{fig3}, the energy gap $E_g$, in the majority spin channel, the indirect band gaps at around Ef along the $\Gamma$-X symmetry, is 0.42062~eV. This gap implies the HM character of compounds and causes $100\%$ spin polarization at Ef. 
The calculated spin-polarized (DOS), total densities of states and atom-projected DOS of the Mn$_2$NbSn Heusler alloy for both  Hg$_2$CuTi and Cu$_2$MnAl structures are presented at their optimized equilibrium lattice constants in figure~\ref{fig2}. The form of total DOS and atom projected DOS of the Mn$_2$NbSn Heusler alloys for two different structures is very different.  For the  Cu$_2$MnAl  type structure, there is a  gap in the majority spin (spin up state), the Fermi level just falls within  the gap in the spin-up band indicating a semiconductor behavior and it crosses the energy bands in the minority spin state which makes Mn$_2$NbSn  Heusler alloys with  Cu$_2$MnAl  type structure  half-metallic (HM) magnetic  compounds at the equilibrium lattice constant. In Hg$_2$CuTi type structure, the Fermi level crosses the energy bands for both majority and minority spin indicating the metallic behavior for this structure. In the majority spin band gap, the valence band maximum is situated at $-0.06572$ and the conduction band minimum is situated at 0.23442.  This energy gap in spin up state leads to $100\%$ spin polarization at the Fermi level, resulting in the half-metallic behavior at the equilibrium state of the Cu$_2$MnAl type structure. For Mn$_2$MoSn, the band in both  Cu$_2$MnAl and Hg$_2$CuTi structures, and for the spin up and the spin down, it is evident that the structure has metallic intersections at the Fermi level, indicating a strong metallic nature of the spin-up and spin-down electrons.


 \begin{table}[!t]
\caption{The calculated values of magnetic moments ($\mu_{\textrm{B}}$) of the of  Mn$_2$YSn (Y = Mo, Nb, Zr) Heusler compounds.}
\label{tbl-smp2}
\vspace{2ex}
\begin{center}
\renewcommand{\arraystretch}{0}
\begin{tabular}{|c|c||c|c|c|c|c|c||}
\hline   
 & &$M_{\text{Mn}_1}$&$M_{\text{Mn}_2}$&M$_\text{Y}$&$M_{\text{Sn}}$&$M_{\text{interstitial}}$&$M_ \text{total}$ ($M_\text{t}$)\strut\\
\hline
\rule{0pt}{2pt}&&&&&\\
\hline
\raisebox{-1.7ex}[0pt][0pt]{Mn$_2$MoSn}
 &  Hg$_2$CuTi&3.473&2.770&$- 0.815$&$- 0.029$&$-0.136$&5.262\strut\\   
\cline{2-8}

& Cu$_2$MnAl&0.004&0.000&$- 0.010$&$- 0.003$&$-0.062$&0.000\strut\\
\hline
\raisebox{-1.7ex}[0pt][0pt]{Mn$_2$NbSn}
 &  Hg$_2$CuTi&3.423&2.539&$- 0.585$&$- 0.052$&$-0.153$&5.170\strut\\
\cline{2-8}
& Cu$_2$MnAl&0.761&0.761&$- 0.331$&$- 0.032$&$-0.165$&1.000\strut\\
\hline
\raisebox{-1.7ex}[0pt][0pt]{Mn$_2$ZrSn}

 &  Hg$_2$CuTi&3.163&2.937&$- 0.182$&$- 0.048$&$-0.061$&5.931\strut\\ 
\cline{2-8}     
& Cu$_2$MnAl&1.262&1.262&$- 0.283$&$- 0.045$&$-0.190$&2.000\strut\\

\hline
\end{tabular}
\renewcommand{\arraystretch}{1}
\end{center}
\end{table}

Total magnetic moment (MT) of  Mn$_2$YSn (Y = Mo, Nb, Zr) full-Heusler alloys, the  atomic moment of each ion and magnetic moment in interstitial zones are calculated at the  equilibrium lattice parameter by  GGA. The obtained results are presented in table~\ref{tbl-smp2}. It can be seen that they are integral values, 1~$\mu_{\textrm{B}}$ for Mn$_2$NbSn and 2$\mu_{\textrm{B}}$ for Mn$_2$ZrSn with Cu$_2$MnAl type structure, and the contributions to the total magnetic moments Mt are mainly attributed to the Mn atom, the Y(Y = Mo, Nb, Zr) and Sn atomic magnetic moments can be neglected. The unequal magnetic moments on the Mn1 and Mn2 atoms, for Hg$_2$CuTi type structure, result from different atomic environments. The negative magnetic moments on the Y (Y= Mo, Nb, Zr), and Sn atoms show that there is antiferromagnetic coupling with the Mn atom. For Mn$_2$MoSn, the total magnetic moment is equal to zero (with Cu$_2$MnAl structure) which confirms the non-magnetic behavior for this compound. 
Magnetic properties can be directly connected to the electronic structure by the Slater-Pauling rule: $ M_\text{t} = Z_t -24$, the $M_\text{t}$ is total spin magnetic moments in the unit cell, and $Z_t$ is the total number of valence electrons. $Z_t$ is equal to 24 for Mn$_2$MoSn $[(7\times2) + 6+4 =24]$, equal to 23 for Mn$_2$NbSn $[(7\times2) + 5+4 =23]$ and it is equal to 22 for Mn$_2$ZrSn $[(7\times2) + 4+4 =22]$.  
In full-Heusler alloys, the minority band comprises 12 electrons per unit cell. Therefore, the 24 valence electrons are equally distributed into both spin directions, and the alloy is nonmagnetic. If the alloy has more than 24 valence electrons, spin polarization will occur and the exchange interaction will shift the majority states to lower energies. The additional electrons will fill in only the majority spins, which results in an integral spin moment \cite{24}.
However, for half-metallic full-Heusler alloys with less than 24 electrons per unit cell, such as our case study, the energy gap is in the majority spin band rather than in the minority spin band. A similar result was found by Anjami et al. \cite{12}.

\section{Conclusion}\label{sec4}

First principles FP -LAPW calculations were performed on Mn$_2$YSn (Y = Mo, Nb, Zr). Based on the results, we predict that Cu$_2$MnAl type structure is more stable than the Hg$_2$CuTi type. It is concluded that the spin polarized band structure and densities of states of the Mn$_2$ZrSn and Mn$_2$NbSn present $100\%$ spin polarization around Fermi level and are half-metallic ferromagnets. The calculated total magnetic moments are 1~$\mu_{\textrm{B}}$ and 2~$\mu_{\textrm{B}}$ for Mn$_2$NbSn and Mn$_2 $ZrSn, respectively, for the Cu$_2$MnAl type structure, which is quite well proved with the Slater-Pauling rule.  

The  Mn$_2$MoSn compound with Cu$_2$MnAl presents a non-magnetic behavior with spin up and spin down symmetric (DOS), the band structures are identical (up and down) and the magnetic moment is equal to 0 $ \mu_{\textrm{B}}$.

\section{Acknowledgement}

This work was supported by DGRST-ALGERIA.

\newpage
\ukrainianpart

\title{Прогнозування електронних та напівметалевих властивостей сплавів Гейслера 
 Mn$_2$YSn (Y = Mo, Nb, Zr) 
}

\author{С. Зеффане\refaddr{label1,label2}, M. Саях\refaddr{label1,label2}, Ф.Дахман\refaddr{label1,label3}, M. Мохтарі\refaddr{label1,label2}, Л. Зекрі \refaddr{label2}, Р. Хената\refaddr{label3}, Н.~Зекрі\refaddr{label2}}
\addresses{
\addr{label1} Інститут природничих наук і технологій, університетський центр Тіссемсілту, 38000 Тіссемсілт, Алжир 
\addr{label2}  Інститут природничих наук і технологій iм.~Мохамеда Будiафа м. Оран, USTO-MB, LEPM, BP 1505,  31000 Оран, Алжир  
\addr{label3} Лабораторія квантової фізики і математичного моделювання (LPQ3M), відділ технологій, університет Маскари, 29000 Маскара, Алжир
 }

\makeukrtitle 
\begin{abstract}
Ми досліджуємо структурні, електронні та магнітні властивості сполук Гейслера M$_2$YSn (Y = Mo, Nb, Zr) з використанням першопринципної теорії функціоналу густини та  узагальненого градієнтного наближення. Встановлено, що розраховані константи гратки добре узгоджуються з теоретичними значеннями. Ми спостерігаємо, що структура типу Cu$_2$MnAl є більш стійкою, ніж структура типу Hg$_2$CuTi. Розраховані сумарні магнітні моменти Mn$_2$NbSn та Mn$_2$ZrSn дорівнюють 1 $\mu_{\textrm{B}} $ і 2 $\mu_{\textrm{B}}$ при  рівноважній сталій гратки 6,18 \AA ~ і 6,31 \AA, відповідно, для структури типу Cu$_2$MnAl. Mn$_2$MoSn  має металевий характер як у структурах типу Hg$_2$CuTi, так і в Cu$_2$MnAl. Повний спіновий магнітний момент підпорядковується правилу Слейтера-Полінга. Напівметал демонструє $ 100 \% $ спінової поляризації на рівні Фермі. Таким чином, ці сплави є перспективними магнітними кандидатами в спінтронних пристроях.

\keywords Гейслер, напівметалевий, магнітний момент, спiнтронний
\end{abstract}
  \lastpage
 \end{document}